\documentclass[useAMS,usenatbib]{mn2e}
\usepackage{graphicx}
\usepackage{amssymb}
\usepackage[authoryear]{natbib}
\usepackage{color}

\bibpunct{(}{)}{;}{a}{}{,}
\title[IFS of two HI rich E+A galaxies]{Integral Field spectroscopy of two HI rich E+A galaxies} 
\author[Michael B. Pracy et al.]{
\parbox[t]{\textwidth}{ Michael B. Pracy$^1$, Matt S. Owers$^2$, Martin Zwaan$^3$, Warrick Couch$^{2}$,  Harald Kuntschner$^{3}$, Scott M. Croom$^{1}$, Elaine M. Sadler$^{1}$}\\
\vspace*{6pt}\\
$^1$Sydney Institute for Astronomy, School of Physics, University of Sydney, NSW 2006, Australia\\
$^2$ Australian Astronomical Observatory, P.O. Box 915, North Ryde, NSW 1670, Australia\\
$^3$European Southern Observatory, Karl-Schwarzschild Strasse 2, 85748, Garching, Germany\\
}

\begin{document}

\date{Received 0000; Accepted 0000}

\pagerange{\pageref{firstpage}--\pageref{lastpage}} \pubyear{2010}

\maketitle

\label{firstpage}

\begin{abstract}
Approximately half of the nearby E+A galaxies followed up with 21-cm observations have detectable HI emission. The optical spectra of these galaxies show strong post-starburst stellar populations but no optical emission lines implying star-formation is not ongoing despite the presence of significant gas reservoirs. We have obtained integral field spectroscopic follow up observations of the two brightest, and nearest, of the six E+A galaxies with HI 21-cm emission in the recent sample of Zwaan et al. (2013). In the central regions of both galaxies the observations are consistent with a post-starburst population with little emission. However, outside the central regions both galaxies have strong optical emission lines, with a clumpy or knot-like distribution, indicating ongoing star-formation.  We conclude that in these two cases the presence of optical spectra lacking evidence for star-formation while a large gas mass is present can be explained by an aperture effect in selecting the nearby E+A galaxies using single-fibre spectroscopy that probes only the galaxy core. 
\end{abstract}


\section{Introduction}
E+A galaxies are identified by their unusual optical spectra: strong Balmer absorption lines signal the presence of a substantial young stellar population ($\lesssim$1\,Gyr) but a lack of optical emission lines indicate little or no current star-formation. Together, these properties suggest these galaxies are in a post-starburst phase of their evolution \citep[e.g.][]{couch87} and may be transitioning from blue star-forming galaxies onto the red sequence. Whether or not the E+A phase represents a final transition from a star-forming to quiescent galaxy will depend on the state of the gas content, since this is the supply for future star-formation.  If the processes causing the E+A phase result in the complete removal or exhaustion of the gas supply then the transformation will be long-lasting. If, however, the truncation of the star-formation is an interim effect as a result of the starburst but a gas reservoir remains, then the E+A phase may be transient with more star-formation to come \citep{zwaan13}.

Only eleven E+A galaxies have measured {\sc HI}-21cm detections corresponding to a detection rate of 50\,per cent \citep{chang01,buyle06,zwaan13}. This detection 
rate is, of course, sensitive to the depth of the surveys and the distances to the galaxies. The small number of detections is in part the result of  
the scarcity of E+A galaxies known in the local universe and amenable to follow-up at 21-cm in reasonable integration 
times.  The E+As detected in {\sc HI} have gas masses intermediate between gas-rich early type galaxies and gas poor spiral galaxies, at corresponding stellar mass \citep{zwaan13}.
This raises the question, for the E+As that have substantial reservoirs of neutral gas which could be used for star-formation, why has the 
star-formation been truncated?  \citet{buyle06} propose that, perhaps, the starburst consumes the dense molecular clouds which ends star-formation 
but a large reserve of more tenuous {\sc HI} remains. 

The only galaxy that \citet{chang01} detected  in {\sc HI} was part of an interacting pair of post-starburst galaxies. 
Higher spatial resolution follow-up of this system by \cite{buyle08} established the majority of the {\sc HI} gas was distributed in surrounding tidal tails and the optical galaxies 
themselves correspond to the lowest column density regions. This removal of the gas from the galaxy, at least temporarily, readily explains the truncation of the 
star-formation. The {\sc HI} detections of \citet{buyle06} \& \citet{zwaan13} were obtained with single dish observations which do not provide 
details about the spatial distribution of the {\sc HI}. For single dish observations the possibility of confusion in the beam needs to be considered. The \citet{buyle06} 
observations using the Parkes Telescope have a primary beam size of $\sim 15$\,arcminutes. Such an area will contain a large number of possible contaminating objects, only 
a small fraction of which can be ruled out using a known redshift. Indeed, \cite{zwaan13} followed up the three E+As in their sample with the highest 21-cm flux densities 
using synthesis imaging observations and found that the {\sc HI} detection of one of the E+As was confused with a nearby spiral galaxy at the same redshift. 

The E+A galaxies that have been detected in {\sc HI} 21\,cm are all at low redshift, ranging from z$\sim 0.005$ to z$\sim 0.1$. Observing galaxies at 
larger distances is difficult owing to the weak flux of the 21\,cm emission line. In addition, E+A galaxies are rare in the local Universe making up a very small fraction of 
the overall galaxy population \citep{blake04,goto07}. As a result, they are generally identified from large spectroscopic surveys such as the Las Campanas Redshift Survey  \citep[LCRS;][]{shectman96}, the Two Degree Field Galaxy Redshift Survey 2dFGRS \citep[2dFGRS;][]{colless01} 
and Sloan Digital Sky Survey  \citep[SDSS;][]{abazajian09}. The spectra from these surveys are obtained using fibres with apertures of 3.5\,arcseconds, 2\,arcseconds and 3\,arcseconds, respectively. For 
nearby galaxies, such as those in the 
sample of \citet{zwaan13}, these apertures do not encompass the entire galaxy but are confined to a small region in the galaxy core. This allows for the possibility that the
post-starburst spectral signature is not a galaxy-wide property and star-formation is ongoing outside the galaxy core.
 
As mentioned above, the three E+As with the highest 21-cm flux densities in the sample of \citet{zwaan13} have been followed up with 
imaging observations using the Jansky Very Large Array (JVLA). One of these E+As was found to be confused with a nearby spiral galaxy; the other two, however, were confirmed to 
have large, well ordered {\sc HI} disks (Klitsch et al. in prep) with total HI masses of $\sim 10^9$M$_\odot$ \citep{zwaan13}. In this paper we present integral field spectroscopy of these two galaxy which provides spatially resolved optical 
spectroscopy over the entire optical extent of the galaxies and removes the aperture bias inherent in the original survey spectra.

\section{The Data}
Our target galaxies are two E+A galaxies from \citet{zwaan13} with confirmed associated {\sc HI} \citep{zwaan13}: ESO534--G001 and 2dFGRS--S833Z022. A summary of the optical and radio
properties of the galaxies is given in Table \ref{tab:targets}. We obtained integral field observations using the Wide Field Spectrograph \cite[WiFeS;][]{dopita07,dopita10},
on the Mount Stromlo and Siding Spring Observatory's 2.3-m telescope. The observations were made on 2012 November 12th in clear conditions with an average seeing of 1.7\,arcseconds. The observations were performed  in Nod and Shuffle mode. The WiFeS field-of-view is 25$\times$38 arcseconds with a spaxel size of 1\,arcsecond. The spectra have a resolution of R$\sim$3000 and span from $\sim 3500$ to $9000$\AA. The data were reduced using the dedicated PyWiFeS data reduction package \citep{childress14}. This results in a fully reduced, coadded and flux calibrated data cube for each galaxy. 
\begin{table*}
\caption{Summary of target properties.}
\begin{center}
\begin{tabular}{llllll}
\hline\hline
NED name               &      RA            & Dec                   &                       z              &  M$_{\rm HI} $             &    b$_J$             \\
                               &      (J2000)      &  (J2000)             &                 &    ($10^8$M$_\odot$) &    (mag)             \\
\hline%
 ESO 534-G001       &   22:36:06.9   &  -26:18:51   &  0.0050    &    9.63              &   15.43              \\ 
 2dFGRS S833Z022  &   04:14:37.4    &     -22:48:25            &  0.0075                      &    14.7              &   16.36              \\
\hline\hline
\end{tabular}
\end{center}
\label{tab:targets}
\end{table*}

\section{Analysis and Results}

\subsection{Spectra}
\setcounter{figure}{0}
\begin{figure*}
\begin{center}
\begin{minipage}{0.95\textwidth}
\includegraphics[scale=1.0,width=0.35\textwidth,angle=90]{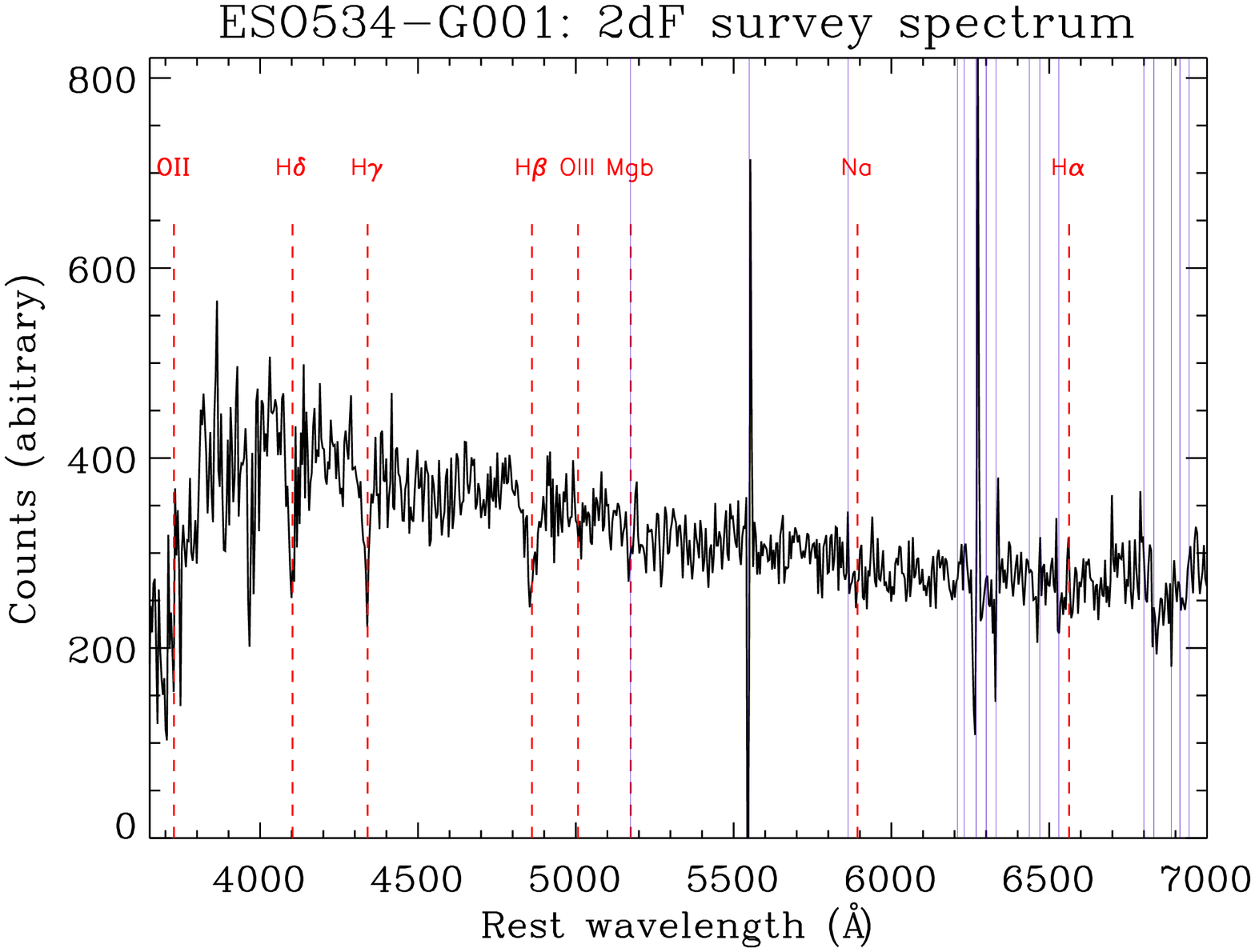} 
\includegraphics[scale=1.0,width=0.35\textwidth,angle=90]{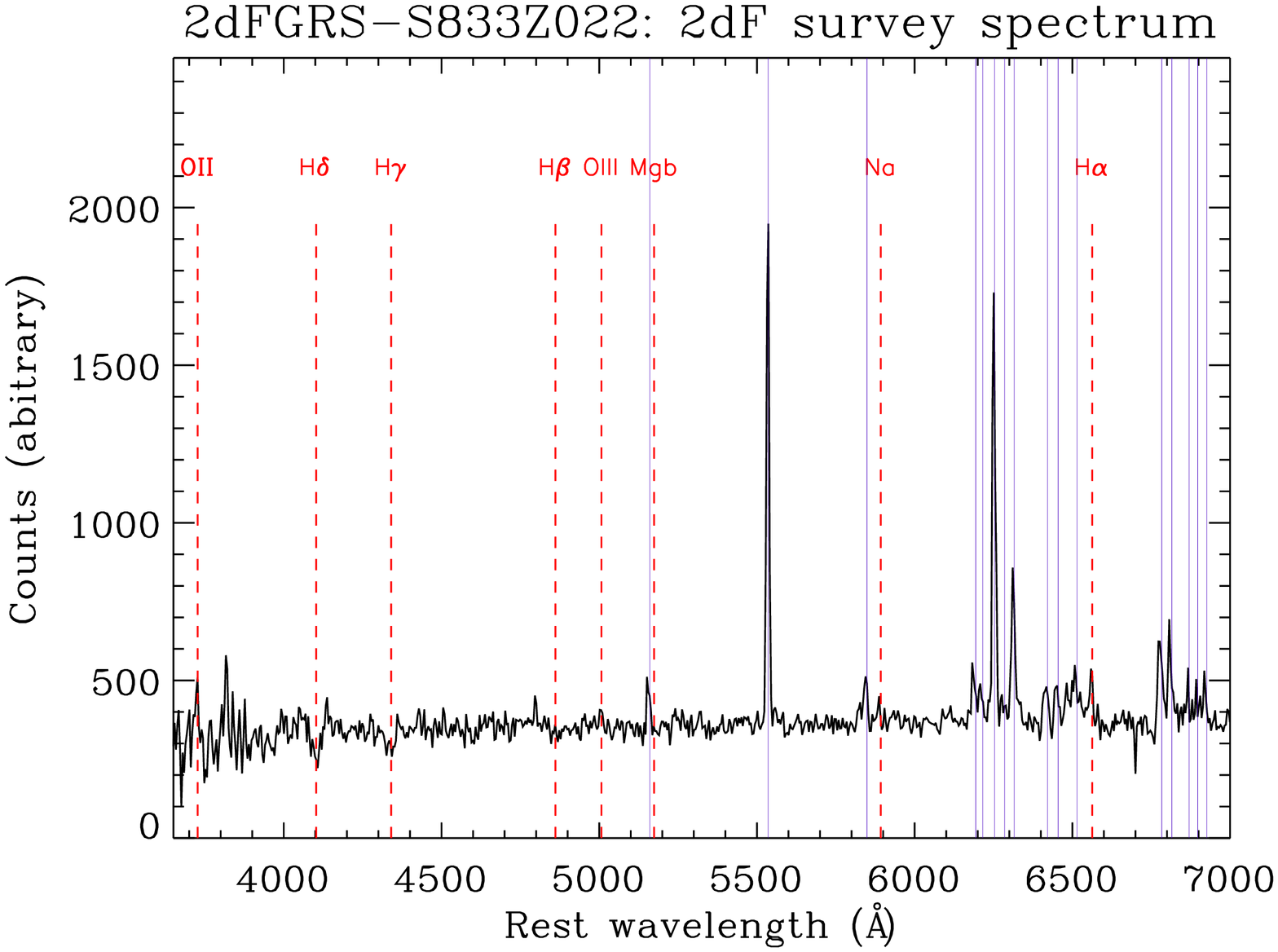} 
\end{minipage}
\begin{minipage}{0.95\textwidth}
\includegraphics[scale=1.0,width=0.35\textwidth,angle=90]{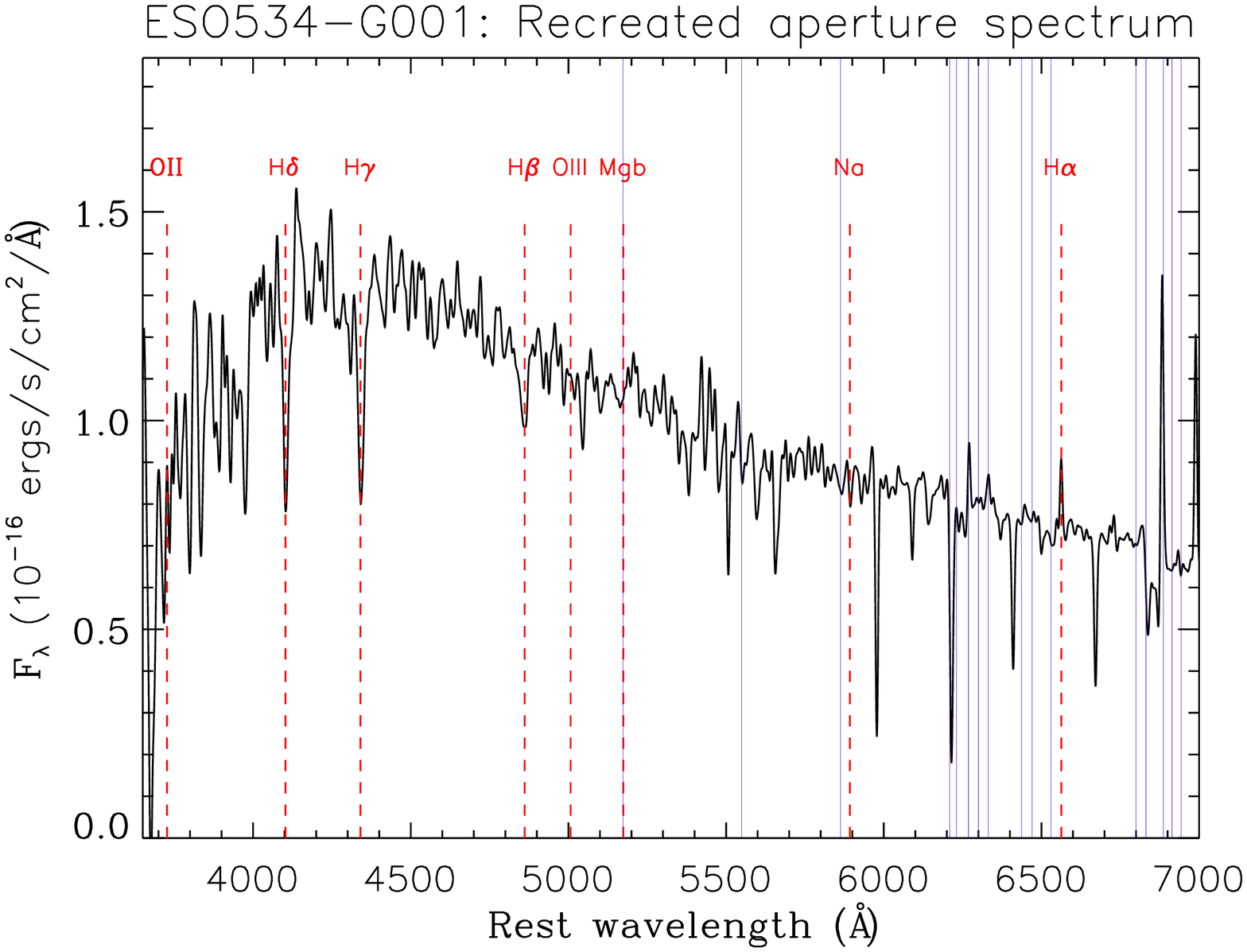} 
\includegraphics[scale=1.0,width=0.35\textwidth,angle=90]{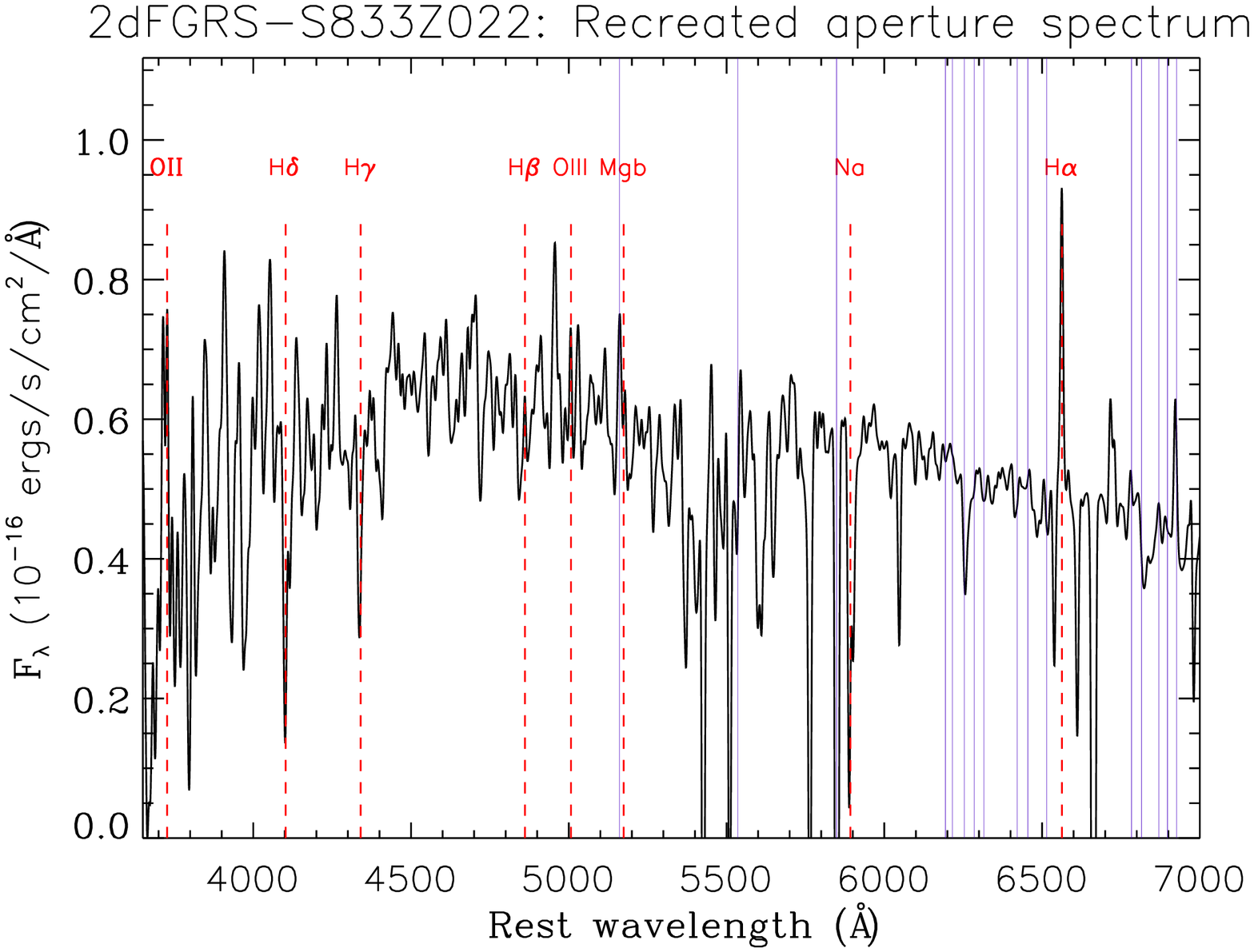} 
\end{minipage}
\begin{minipage}{0.95\textwidth}
\includegraphics[scale=1.0,width=0.35\textwidth,angle=90]{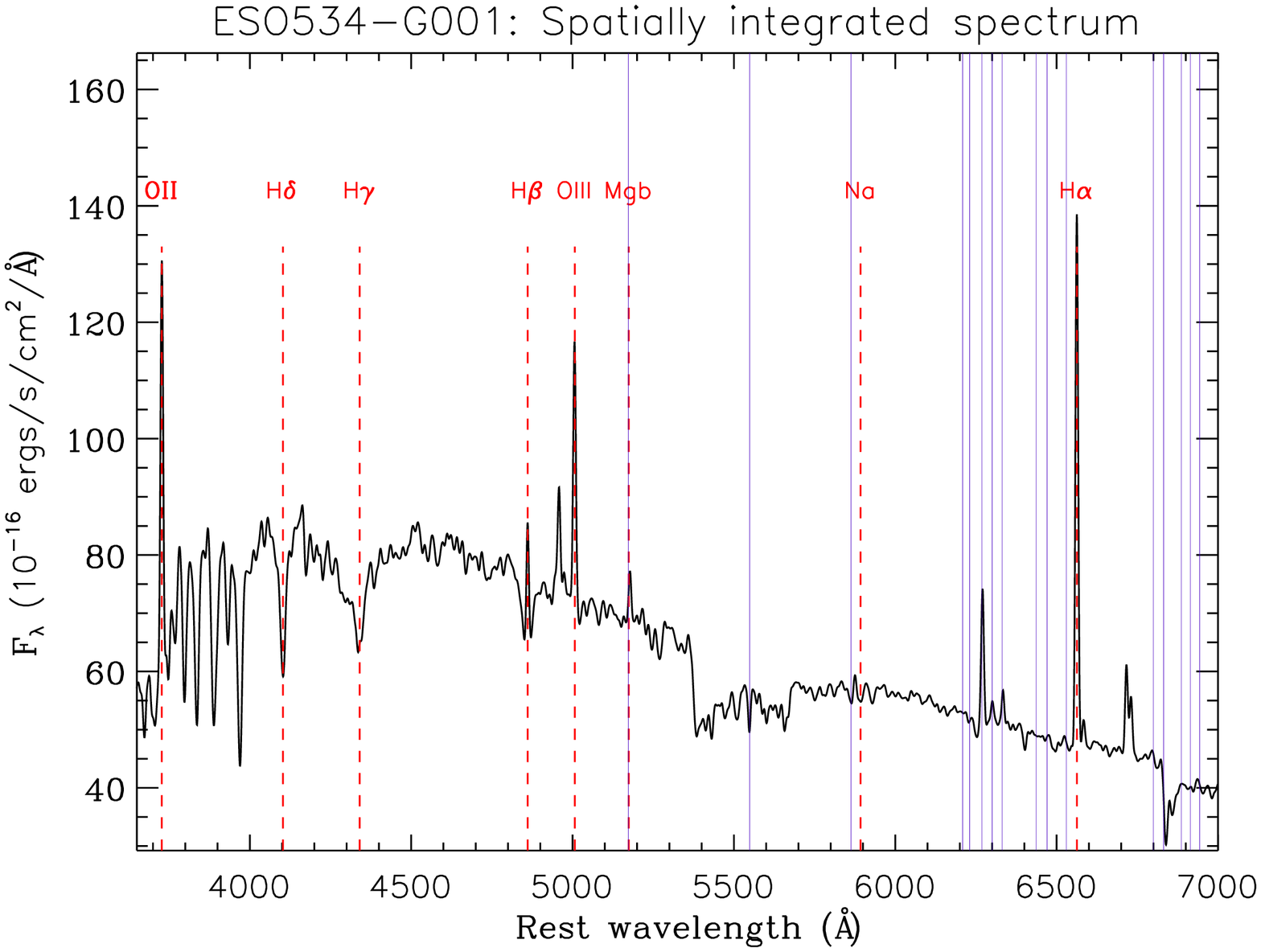} 
\includegraphics[scale=1.0,width=0.35\textwidth,angle=90]{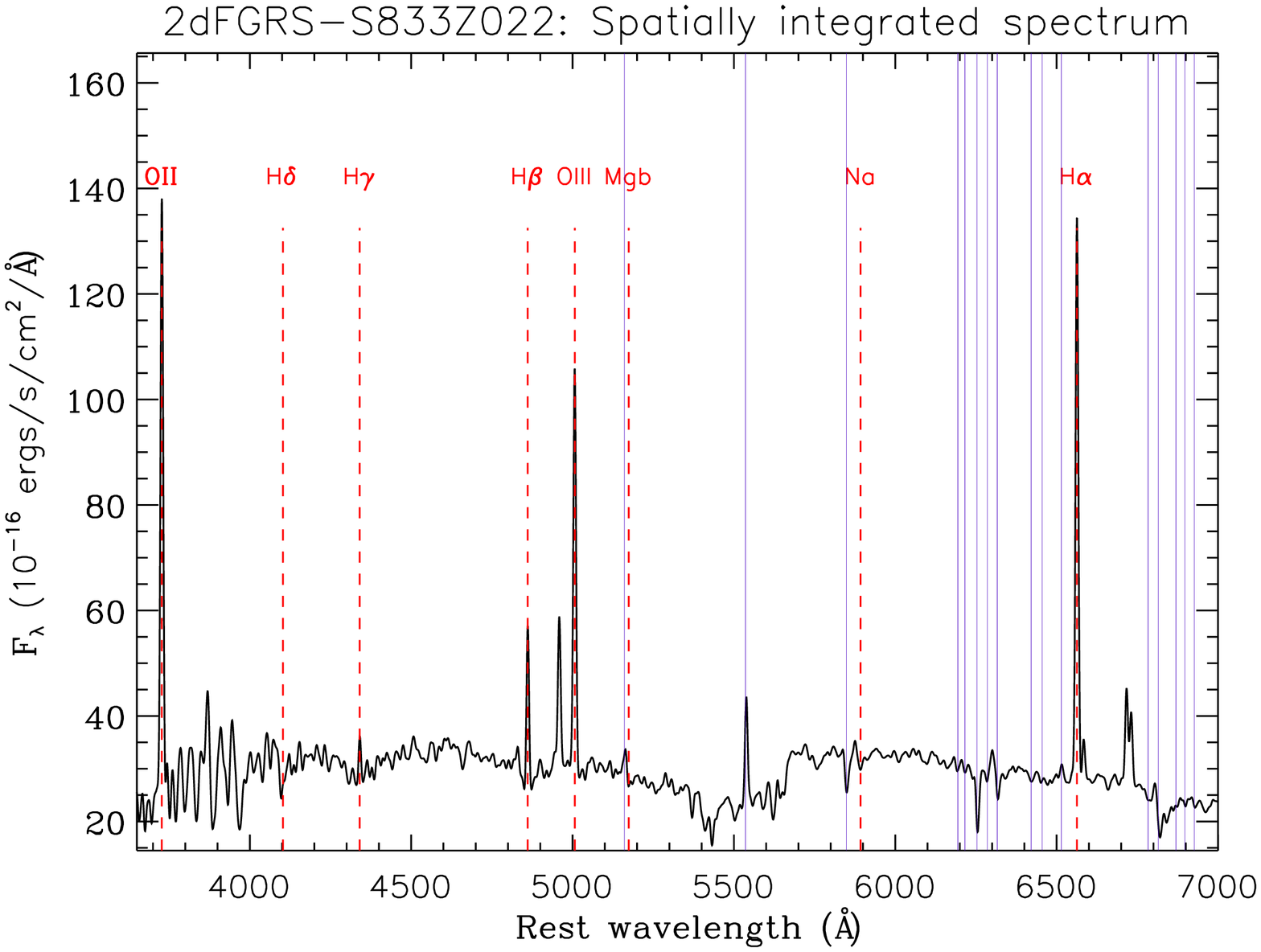} 
\end{minipage}
\end{center}
\vspace{-0.3cm}
\caption{The left column shows spectra for ESO534-G001 and the right column shows spectra for
2dFGRS-S833Z022. {\it Top row:} The 2dFGRS spectra from which the galaxies were classified as
E+A galaxies.  The spectra were obtained with a central 2 arcsecond aperture
fibre. Some of the prominent spectral features are marked with the
red dashed lines and the positions of sky lines marked by blue
vertical lines.  Both spectra show strong Balmer absorption lines
and little emission. {\it Middle row:} 2 arcsecond aperture spectra reconstructed from our integral field spectroscopy. For ESO534-G001 the aperture is centred on
the peak of continuum emission. In the case of 2dFGRS-S833Z022 it has been slightly offset to try and best match the 2dFGRS spectrum. {\it Bottom row:} Integrated spectra constructed by coadding the entire data cubes. Our spectra ({\it middle and bottom rows}) have been smoothed with a Gaussian of $\sigma=4$\AA\, for presentation.}
\label{fig:spectra}
\end{figure*}

E+A galaxies are typically selected and defined based on the equivalent widths of the H$\delta$ and [OII]$\lambda 3727$ lines \citep[e.g.][]{zabludoff96,dressler99,blake04,goto07}. The precise definitions vary between authors but the upper-limit on the strength of the [OII]$\lambda 3727$ is generally: 
$$\rm{EW([OII]} \lambda 3727) > [-2.5,-5]\rm{\AA}$$ 
Note, we have used (and will throughout) the standard convention that emission line equivalent widths are negative and absorption line equivalent widths are positive. The classifications can be made more robust by using a combination of the Balmer absorption lines \citep{zabludoff96,blake04} and, more critically, the H$\alpha$ line as a constraint on emission \citep{blake04,goto07}. The H$\alpha$ line is a superior constraint since it is intrinsically stronger, more simply related to the star-formation rate and less affected by dust than the [OII]$\lambda 3727$ line. However, the H$\alpha$ line is redshifted out of optical spectra at moderate redshift (z$\sim 0.4$) and at low redshift can be contaminated by night sky emission lines and telluric absorption lines. 

The {\it top} row of Figure \ref{fig:spectra}  shows the 2dFGRS spectra of our two target galaxies from which they were classified as E+A galaxies. 
They have strong H$\delta$ absorption and little or no [OII]$\lambda 3727$ emission. Both galaxies have weak H$\alpha$ emission, although, the emission does not stand out from the nearby night sky residuals. 
While the 2dFGRS spectra are not flux calibrated we have applied the average response curve of \cite{lewis02} so the spectra have approximately the correct spectral shape.
In the {\it middle} row of Figure \ref{fig:spectra} we present spectra constructed from our integral field observations with an aperture matched in size to the original 2dFGRS spectra i.e. 2 arcseconds in diameter. In the case of ESO534-G001 (left hand side) the matched aperture spectrum is extracted from the position corresponding to the peak in the continuum flux. In the case of 2dFGRS-S833Z022 (right hand side) the peak in the continuum flux corresponds to a region with strong emission lines. Comparison of the astrometry from the 2dFGRS catalogue \citep{colless03} and Digital Sky Survey images of the galaxy indicate that the placement of the fibre was off centre -- likely as a result of a second peak in the flux shifting the centroid. The positions of our matched apertures are shown in Figure \ref{fig:maps} as {\it red circles}. The reconstructed aperture spectra also have strong absorption lines and weak emission. The H$\alpha$ line is seen more prominently as a result of the higher resolution spectra (less smeared out) and the better sky subtraction produced by our nod and shuffle observations. 
The equivalent widths of the H$\alpha$ lines in these spectra are -2.1\AA\, and -10.7\AA\, for ESO534-G001 and 2dFGRS-S833Z022, respectively.
The {\it bottom} row shows the spectra produced by summing the spectra over the entire field-of-view. These spatially integrated spectra have strong emission lines with line ratios consistent with star-formation. 

This explains the dual observations of a post-starburst spectrum with little evidence for star-formation and a high column density HI disk. The E+A signature is confined to the central part of the galaxy where star-formation has been truncated but the large overall HI-mass can still be fuelling star-formation in other regions. In these cases at least the combination of an E+A optical spectrum and a large HI mass can be explained by an aperture effect caused by the fibre aperture subtending only a small fraction of the observed galaxy.

\subsection{Line maps}
In Figure \ref{fig:maps} we use our integral field spectroscopy to spatially map the equivalent widths or line flux of some of the key spectral features. The original E+A classification can be seen in these maps.
For ESO534-G001 (top row) the centre of the galaxy has an  [OII]$\lambda 3727$ equivalent width consistent with zero and  H$\delta$ absorption stronger than 5\AA. For 2dFGRS-S833Z022 (bottom row) there is emission coinciding with the peak of the continuum flux but there is a region slightly offset (to the right-hand side in Figure 2)  with little emission. This region also has strong H$\delta$ absorption and coincides with the  position at which the original 2dFGRS spectrum was taken. The right-most column in Figure \ref{fig:maps} shows the H$\alpha$ flux maps. They are broadly consistent with the [OII]$\lambda 3727$ maps and demonstrate there is significant star-formation ongoing in the outer regions of these galaxies.

The maps in Figure \ref{fig:maps} do not show uniform  emission outside the galaxy core but instead clumps or knots of strong emission. Each galaxy has a few such knots surrounded by more tenuous emission -- this is seen most clearly in the H$\alpha$ flux maps in the right-most column of Figure \ref{fig:maps}.  We can estimate the size of these clumps by comparing their angular extent, via a Gaussian fit to the H$\alpha$ emission, to the point spread function (PSF). Our observations do not give us a simultaneous measurement of the seeing -- which dominates the PSF. However measurements of the seeing were recorded periodically throughout the night of the observations using stars in the acquisition images. The seeing varied between 1.5--2.2 arcseconds, with both targets observed in an average seeing (based on measurements straddling the observations of that target) of 1.7\,arcseconds. These measurements are  neither simultaneous or take into account any degradation in the point spread function from the nod-and-shuffle process but are the best estimate we can make of the PSF in our integral-field-observations. Based on this, the three brightest H$\alpha$ clumps in each galaxy are  resolved with
measured FWHM of 4.0--6.6\,arcseconds corresponding to
 physical size FWHM (assuming  a 1.7\,arcsecond Gaussian PSF) in the range 400--700\,pc. These sizes should be consider upper-limits since we do not know our PSF well and most processes will increase the size of the true PSF, such as inaccuracies in the dithering during nod-and shuffle and intermittent loss of tracking (of which there were several). Taken at face value these sizes are at the high end of the size distribution for extra-galactic HII regions \citep{hunt09}.

\setcounter{figure}{1}
\begin{figure*}
\begin{center}
\begin{minipage}{0.95\textwidth}
\includegraphics[scale=1.0,width=0.32\textwidth,angle=90]{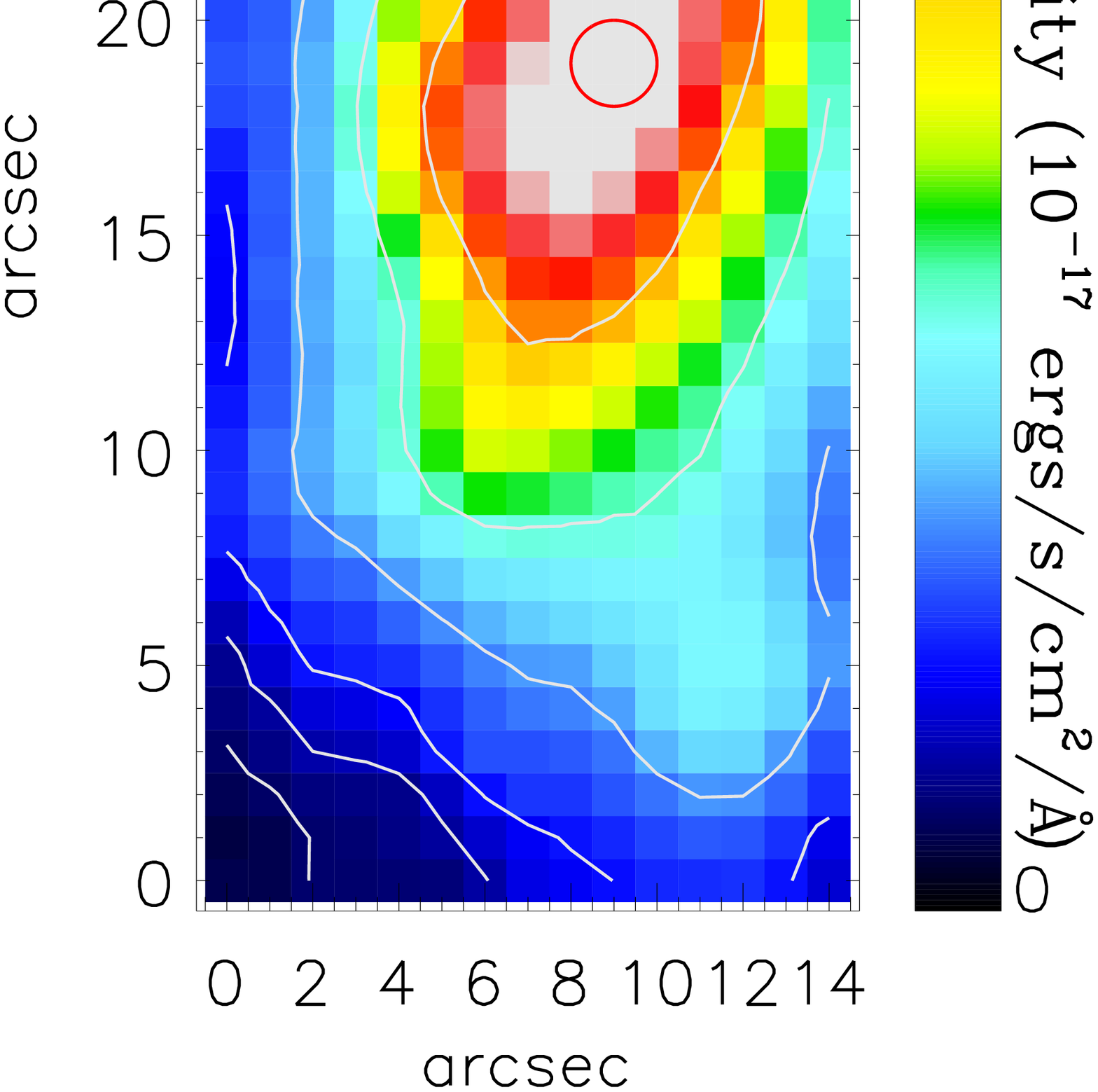} 
\hspace{-0.5cm}
\includegraphics[scale=1.0,width=0.32\textwidth,angle=90]{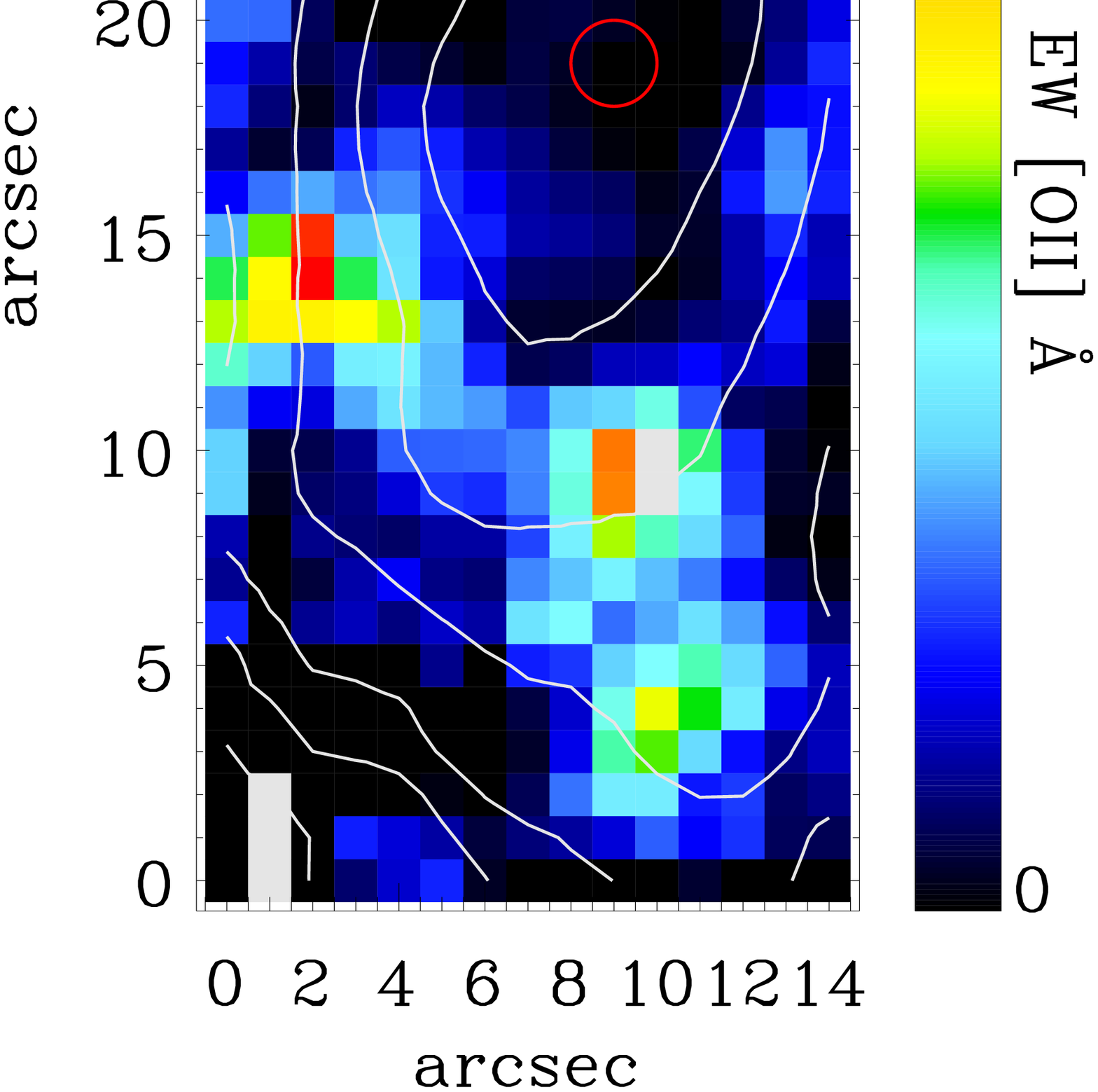} 
\hspace{-0.5cm}
\includegraphics[scale=1.0,width=0.32\textwidth,angle=90]{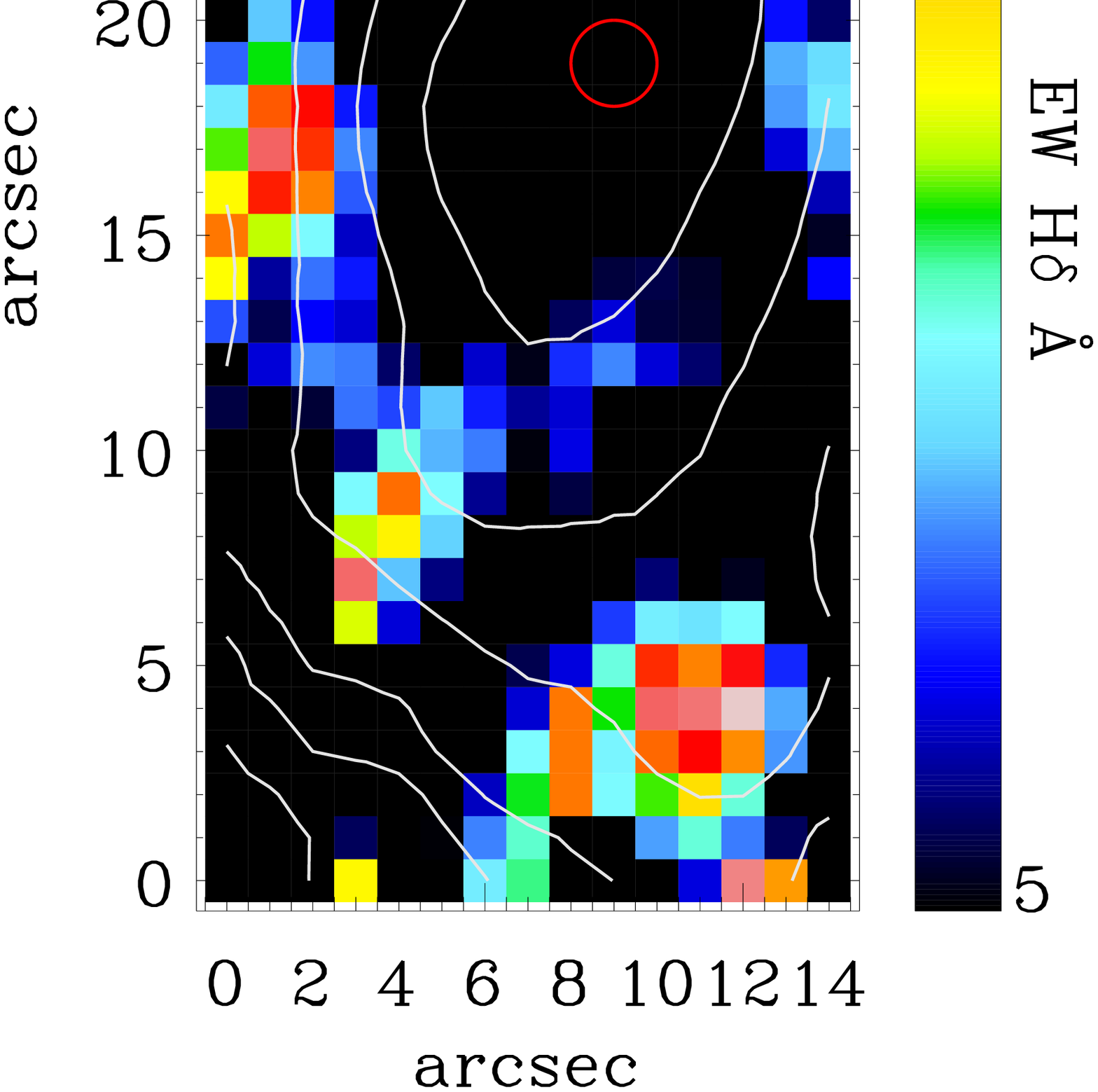} 
\hspace{-0.5cm}
\includegraphics[scale=1.0,width=0.32\textwidth,angle=90]{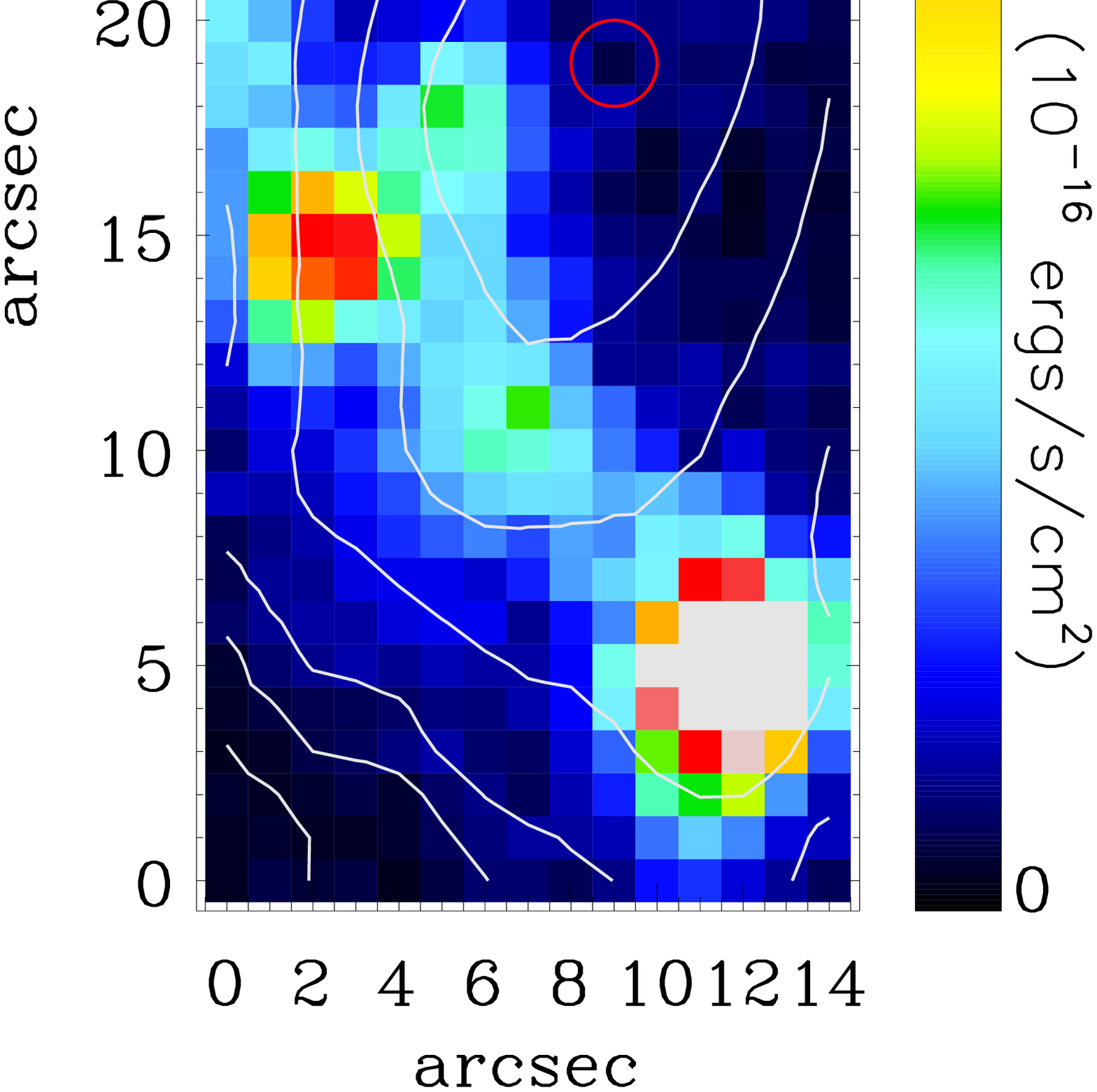} 
\end{minipage}
\begin{minipage}{0.95\textwidth}
\includegraphics[scale=1.0,width=0.32\textwidth,angle=90]{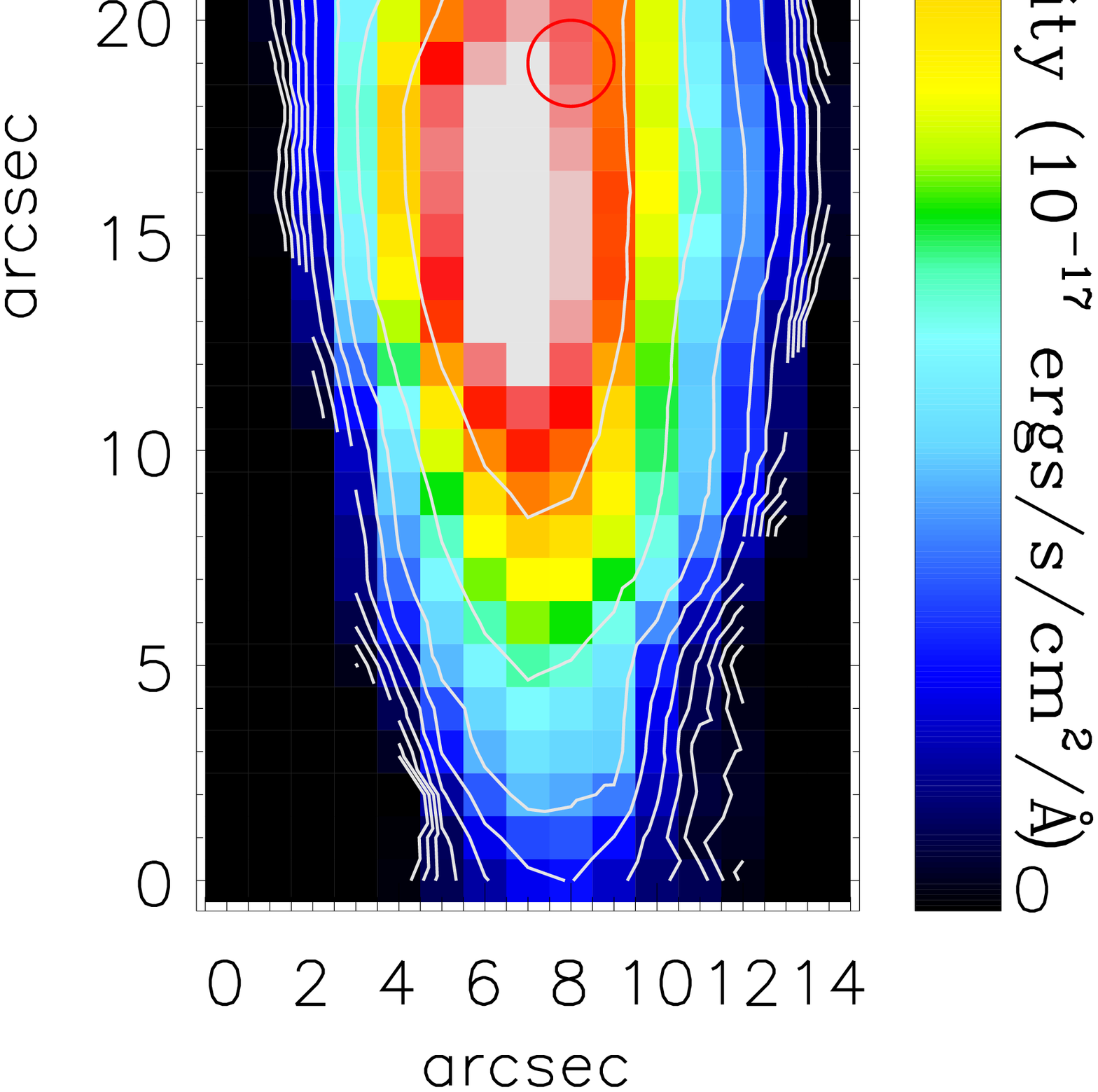} 
\hspace{-0.5cm}
\includegraphics[scale=1.0,width=0.32\textwidth,angle=90]{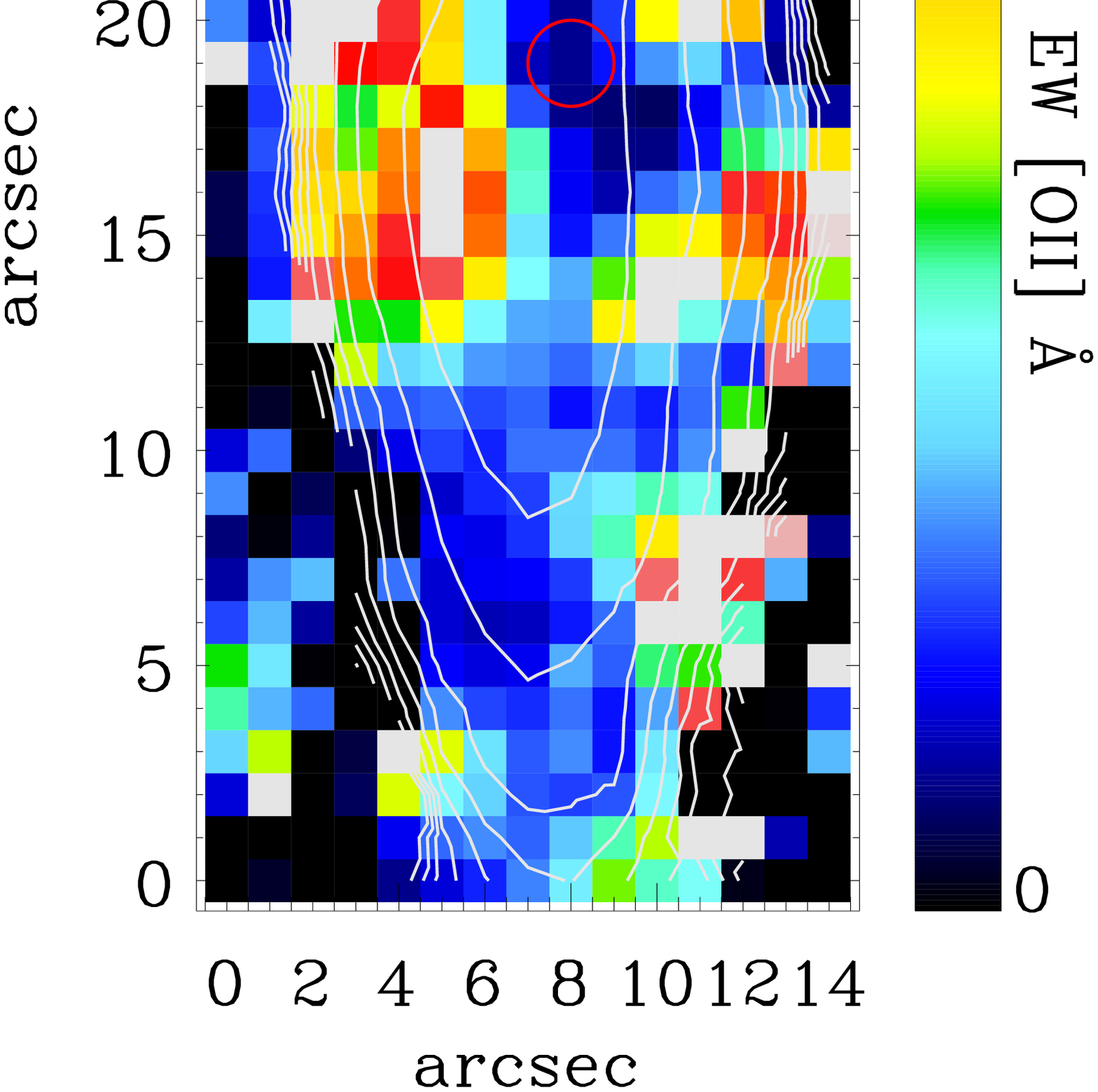} 
\hspace{-0.5cm}
\includegraphics[scale=1.0,width=0.32\textwidth,angle=90]{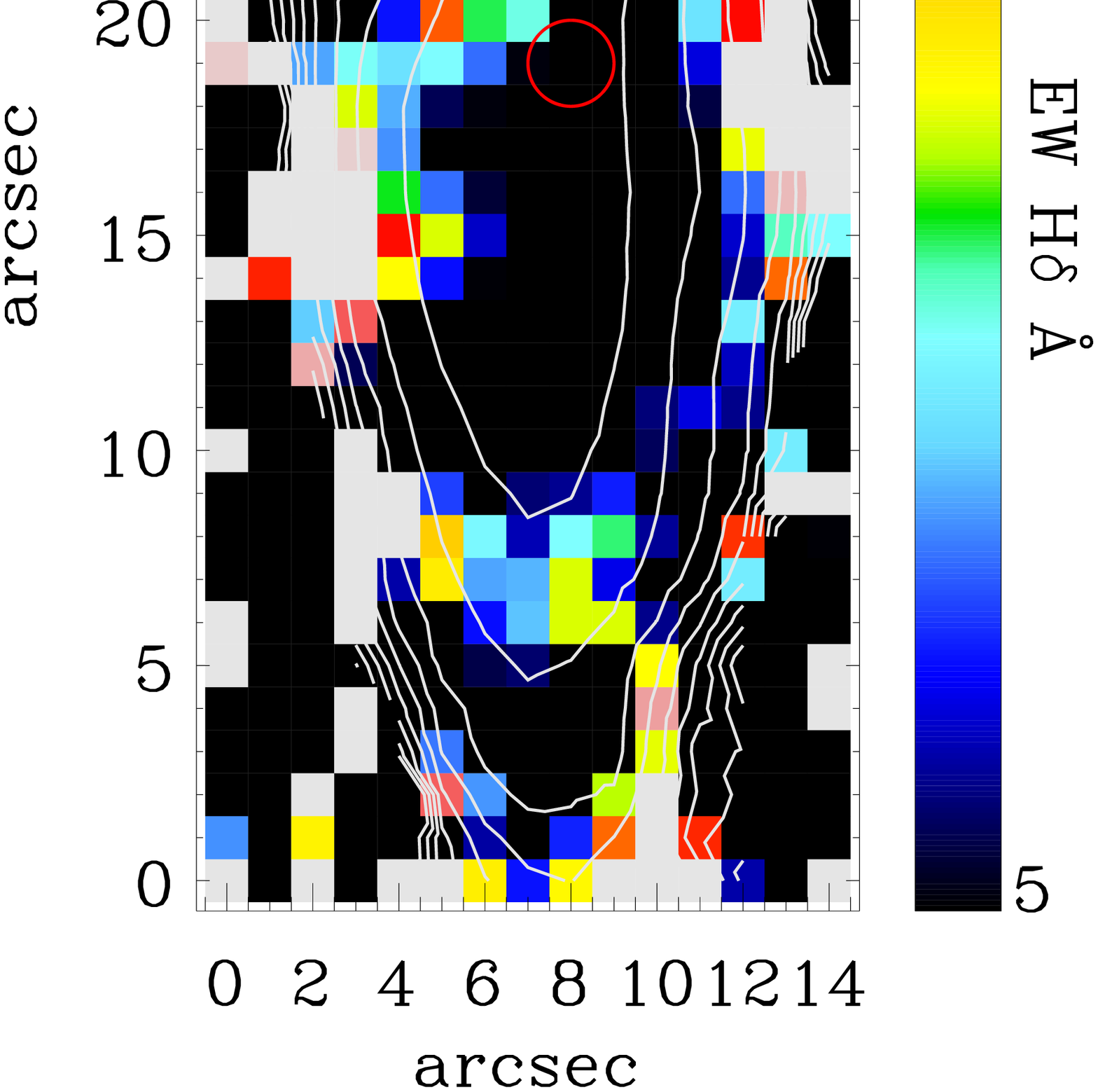} 
\hspace{-0.5cm}
\includegraphics[scale=1.0,width=0.32\textwidth,angle=90]{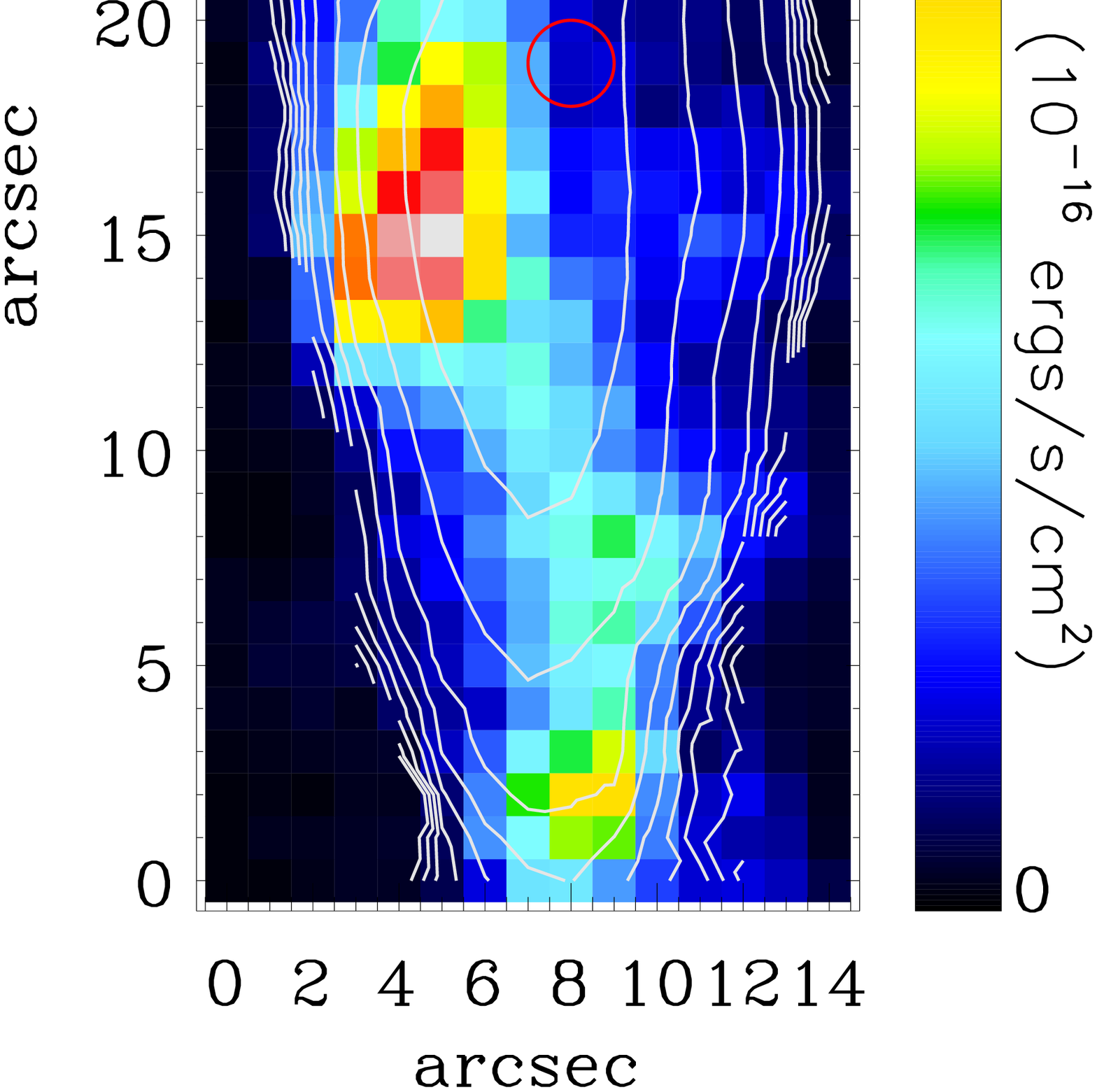} 
\end{minipage}
\end{center}
\vspace{-0.3cm}
\caption{The top row and bottom row show maps for ESO534-G001 and 2dFGRS-S833Z022, respectively.  From {\it left to right:} Continuum flux density (measured using the median flux density between 4000\AA\, and 5000\AA), equivalent width [0II]$\lambda 3727$, equivalent width H$\delta$, and equivalent width H$\alpha$. The white contours trace the continuum flux. The position angle of the long axis of the IFU is  $345^{\circ}$ for ESO534-G001 and $335^{\circ}$ for 2dFGRS-S833Z022. The red circle indicates the size and position of the aperture used to construct the matched aperture spectra in the middle row of Figure \ref{fig:spectra}.}
\label{fig:maps}
\end{figure*}

\section{Discussion and conclusions}
Using integral field spectroscopic observations of two nearby HI-rich E+A galaxies we have demonstrated the coincident observations of a post-starburst, emission line free spectrum and a high column density of HI can be explained by an aperture bias affecting the original optical spectra. In these cases, this aperture effect solves the problem of having a large cold gas reservoir available for star-formation and an optical spectrum inconsistent with significant ongoing star-formation. However, such an effect cannot be the explanation of all the E+A galaxies detected at 21-cms. The most distant of the two objects discussed here is at a redshift of z$\sim 0.0075$. At this redshift the 2 arcsecond fibre used to acquire the original spectrum corresponds to just $\sim 0.3$\,kpc. In contrast, the nearest of the HI detections in \citet{buyle06} is at a redshift $z\sim 0.06$ implying the original LCRS fibre subtends a physical size of $\sim 4$\,kpc. This size is typical of the scale length of a massive galaxy and means the original fibre covered the majority of the source and should have minimal aperture bias. It is also not the case that nearby samples of E+A galaxies, in general, contain a high fraction of star-forming objects misclassified as the result of aperture effects \citep{pracy12,pracy13}.

While the E+A galaxies in the \citet{buyle06} sample should not suffer from aperture bias in the optical, confusion in the radio observations remains a possibility. The single dish observations have a primary beam of $\sim$15 arcminutes and contain a large number of potential interlopers. Indeed, one of the E+A galaxies in the \citet{zwaan13} sample followed up with 21-cm imaging observations had been confused with a nearby spiral galaxy.

The one E+A galaxy detected by \citet{chang01} was at a similar redshift to the \citet{buyle06} objects (i.e. z=0.075) and should not suffer from significant aperture effects in the optical. Synthesis imaging at 21-cm demonstrated that the majority of the gas in this object has been removed by the ongoing interaction with a companion galaxy. Other than this object none of the other HI detected E+A galaxies are free from the possibility of aperture effects in the optical, such as the two objects discussed in this paper, or confusion in the 21-cm detections, such as the higher redshift objects in both the \citet{zwaan13} and \citet{buyle06} samples. The three E+A galaxies detected in HI which have spatially resolved radio observations and spatially integrated optical observations all have straight-forward explanations for the presence of an E+A spectral signature and high HI column density. In the case of  \citet{chang01} object it is the removal of the HI gas in a tidal interaction and for the two objects discussed here it is the result of an aperture bias in the optical observations. The confirmation of true E+A galaxies, devoid of star-formation, coincident with large amounts of cold gas requires both spatially integrated optical observations and spatially resolved 21-cm observations.  Identifying objects from multi-object integral field unit surveys, such as the Sydney-AAO Multi-object Integral Field spectrograph \citep[SAMI;][]{croom12} galaxy survey, which provide multiple spectra across the extent of the galaxy from the outset would alleviate such bias in the optical selection.

What evolutionary scenarios could result in a galaxy with a large column of HI, a post-starburst region near the centre and large clumps of star formation in the outer regions? One possibility is a scheme similar to the one proposed by \citet{buyle06} where the molecular clouds are consumed in the starburst but more tenuous HI remains. In this case the consumption of the molecular gas is only complete in the galaxy centre, where faster timescales for gas consumption are expected \citep{kennicutt98}. Another possibility is that the clumpy star-formation is the product of feedback from the star-burst in the centre.  Many processes which induce starburst-like activity in galaxies such as tidal interactions and mergers result in centrally concentrated star formation. Tidal torques transport angular momentum from the gas to the stars \citep[e.g.][]{barnes96} which moves the gas toward the galaxy centre and can induce a compact central starburst \citep{mihos94,mihos96,barnes96,hopkins09}. The feedback from this starburst from the combination of stellar winds and supernovae can form new molecular gas further out by sweeping up and compressing the atomic medium \citep{dawson13}. In this scenario we expect a young stellar population in the centre surrounded by a HI shell and clumpy HII regions \citep[see e.g.  Figure 1 of][]{dawson13}. Differentiating between these formation scenarios and understanding the physical mechanism responsible for the state of these galaxies will  require higher resolution HI imaging and molecular line observations. We have recently obtained higher resolution HI imaging using the JVLA and CO(1-0) observations with ALMA, the analysis of which will be presented in a forthcoming paper (Klitsch et al. in prep).

\section{Acknowledgments}
This research has made use of the NASA/IPAC Extragalactic Database (NED) which is operated by the Jet Propulsion Laboratory, California Institute of Technology, under contract with the National Aeronautics and Space Administration. M.S.O. acknowledges the funding support from the Australian Research Council through a Super Science Fellowship (ARC FS110200023). SMC acknowledges the support of an ARC future fellowship (FT100100457).

%%%%%%%%

\bsp

\bibliographystyle{mn2e}
\bibliography{references}

\label{lastpage}

\end{document}